\documentclass[conference]{IEEEtran}
\usepackage{cite}
\ifCLASSINFOpdf
   \usepackage[pdftex]{graphicx}
\else
   \usepackage[dvips]{graphicx}
\fi

\ifCLASSOPTIONcompsoc
\usepackage[caption=false,font=normalsize,labelfon
t=sf,textfont=sf]{subfig}
\else
\usepackage[caption=false,font=footnotesize]{subfi
g}
\fi

\usepackage{amsmath}
\usepackage{array}
\usepackage{url}
\usepackage{comment}

\usepackage{color}

\usepackage{xcolor,colortbl}

\begin{document}
\title{Are Donation Badges Appealing? A Case Study of Developer Responses to Eclipse Bug Reports}
%\title{Assessment of Donation Badges: Causal Inference for a Study of Eclipse Bug Repository}

\begin{comment}
\author{\IEEEauthorblockN{Hideaki Hata}
\IEEEauthorblockA{School of Electrical and\\Computer Engineering\\
Georgia Institute of Technology\\
Atlanta, Georgia 30332--0250\\
Email: http://www.michaelshell.org/contact.html}
\and
\IEEEauthorblockN{Homer Simpson}
\IEEEauthorblockA{Twentieth Century Fox\\
Springfield, USA\\
Email: homer@thesimpsons.com}
\and
\IEEEauthorblockN{James Kirk\\ and Montgomery Scott}
\IEEEauthorblockA{Starfleet Academy\\
San Francisco, California 96678--2391\\
Telephone: (800) 555--1212\\
Fax: (888) 555--1212}}
\end{comment}


\author{\IEEEauthorblockN{Keitaro Nakasai\IEEEauthorrefmark{1},
Hideaki Hata\IEEEauthorrefmark{1},
Kenichi Matsumoto\IEEEauthorrefmark{1}}
\IEEEauthorblockA{\IEEEauthorrefmark{1}Graduate School of Information Science,
Nara Institute of Science and Technology, Japan\\ \{nakasai.keitaro.nc8, hata, matumoto\}@is.naist.jp}}


\maketitle

\begin{abstract}
%Managing an effective donation program has become crucial for sustainable open source software projects.
%These projects commonly offer several benefits to attract donations.
Eclipse, an open source software project, acknowledges its donors by presenting donation badges in its issue tracking system Bugzilla.
However, the rewarding effect of this strategy is currently unknown.
We applied a framework of causal inference to investigate relative promptness of developer response to bug reports with donation badges compared with bug reports without the badges, and
estimated that donation badges decreases developer response time by a median time of about two hours.
The appearance of donation badges is appealing for both donors and organizers because of its practical, rewarding and yet inexpensive effect.
\end{abstract}

% \fi
\IEEEpeerreviewmaketitle

\section{Introduction}

%Many Open Source Software (OSS) projects are collecting donations to continue to operate their projects.
%The Eclipse Foundation is one of OSS projects collecting donations.
%It has provided some benefits for its donor.
%\todo{Donations in OSS, in general}
Donations play an important role in open source software (OSS) projects.
%Such financial supports are used to expand infrastructure, hire developers, and support developer communities.
% https://www.reddit.com/r/libreoffice/duplicates/6zf7qc/how_donations_to_the_document_foundation_in_2016/
% https://www.linuxfoundation.jp/about/linux-donate
% https://www.eclipse.org/donate/faq.php
LibreOffice, an OSS project, reported in 2016 that they had received 200,000 donations in three years, and emphasized that a large open source software project does not need a single large corporate sponsor, so long as it can rely on a large and diverse ecosystem of OSS community~\cite{libreofficeblog2016}.
This makes the management of effective donation programs all the more important to maintain sustainable OSS projects.

Despite the criticality of donations in OSS, very few studies have been undertaken on monetary donations to OSS projects, paying much less attention on their effect.
The factors that impact donations were investigated with public records of SourceForge, with a result that a decision to donate was influenced by relational commitment with the OSS platform~\cite{KRISHNAMURTHY2009404}.
Another study identified the composition of the donor groups and the committer group, although small in number, committers donated more than other of the groups~\cite{7925419}.
%The practical rewarding effect of the donations on donors is unknown.

%Currently, in the OSS project Eclipse, the badges appear in Bugzilla, its issue tracking system, to recognize the contributions made by donors, and are held as a form of reward.
Eclipse started its donation program \textit{Friends of Eclipse} in December 2007.
%Their donation program has undergone a series of improvements,
%adding benefits to the donors inspired by the Linux donation program~\cite{bug282088}.
%At its peek in October 2014, the program offered several benefits to donors who contributed 100 USD.
%(See Figure~\ref{fig:benefits2014}).
Showing badges started in November 2014 on the Bugzilla issue tracking system~\cite{bug434249}.
%Although some items, such as T-shirts and a mirror site, have been removed from the benefits, donation badges have remained even after a skepticism raised by a marketing specialist as part of a recent bug report in April 2017~\cite{bug514954} against providing more benefits.
%based on an Eclipse survey result that showed that the donors' relative disinterest in material benefits (Table~\ref{tab:survey}).
Currently, donors who contribute 35 USD or more qualify for the Friend of Eclipse status for one year, and %they can use the Friend of Eclipse logo. These donors 
are recognized on Bugzilla issue tracking system of a friend of Eclipse badge (hereafter called \textit{donation badges}).

However, little is known about the impacts of donation badges. In particular, %how the presence of the donation badges affects the operation of the OSS project and 
how badges might benefit the donors.
Based on the framework for causal inference~\cite{Imbens:2015:CIS:2764565}, we study how promptly developers respond to bug reports that have donation badges compared with bug reports without donation badges.
The analysis revealed that the donation badges decreased response time by about two hours in median.
Our findings suggest that the appearance of donation badges has a practical rewarding effect for individual donors. 
We theorize that this behavior can be explained as an effective and inexpensive \textit{signaling} system, where developers use technical and social information as signals to evaluate potential contributions \cite{Tsay:2014:IST:2568225.2568315}.
We believe that other OSS organizers are able to adopt this strategy to manage their developer ecosystems.

%\begin{figure}[t]
%\centering
%  \includegraphics[width=0.8\linewidth]{fig/benefits2014.jpeg}
%  \caption{A campaign for the Eclipse donation program advertised in Eclipse Member Newsletter issued on October 7, 2014. Online: https://www.eclipse.org/community/newsletter/2014/2014October.html}
%  \label{fig:benefits2014}
%\end{figure}

%\begin{table}[!t]
%\caption{A survey conducted by the Eclipse Foundation in June 2016~\cite{bug514954}. About 80\% of respondents had been using Eclipse for over five years.}
%\label{tab:survey}
%\centering
%\begin{tabular}{lr}
%\hline
%I want to give back since Eclipse technology is free & 84.4\% \\
%I want to support open source & 71.9\% \\
%I want a t-shirt/ other benefits & 5.9\% \\
%I want to help improve Eclipse technology & 72.6\% \\
%I feel a sense of closeness to a community or group & 9.6\% \\
%\hline
%\end{tabular}
%\end{table}

%The donation badge is identified donor in the bug tracking system of Eclipse.
%The donor who donated over US\$35 per year can use the donation badge. 

%In this case study,We focus on the donation badge which is one of the benefits on the Eclipse Foundation.

%Here is refference of donation's study and Eclipse Foundation's approach in donation.
%Write our study and other donation's study. 

\section{Causal Inference in Brief}
\label{sec:ci}
%Our study design applied causal inference techniques.
%This section describes causal inference techniques which use this study.
%First, We apply propensity score matching to reporters data because all reporters data has biased people. After propensity score matching, the reporters has comparable people. Second, we perform quantile difference in differences for the reports.

Causal inference stems from social science that explores cause and effects as its main concern~\cite{angrist2008mostly}.
In Econometrics differences-in-differences methods are one of key analytical elements for causal inference~\cite{angrist2008mostly}.
We adopted this element %and propensity score matching
in our analysis, as outlined the figure below.
Differences in differences (DID) statistically visualizes actual and counterfactual scenarios, thereby enabling a causality analysis.
For the inquiry of the effects of a treatment in statistics, one cannot see both results with and without a treatment based on one individual only.
DID addresses this problem by comparing two groups, one with a treatment and one without it.
% remove subsection
%\subsection{Differences in Differences}

%Something -> Employmentに変更
\begin{figure}[t]
\centering
  \includegraphics[width=\linewidth]{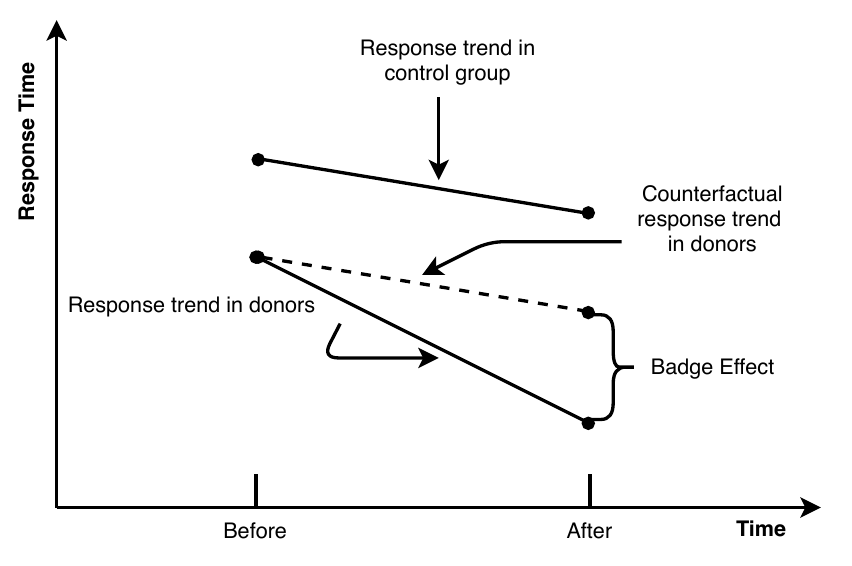}
  \caption{Example of causal inference framework using a DID model (response time vs. before and after donation badge introduction)}
  \label{fig:DIDmodel}
\end{figure}
Figure~\ref{fig:DIDmodel} shows how DID is used to understand the effect of donation badges.
We illustrate the response times of two groups at the period of before and after the donation badge program was introduced.
Donors refer to all contributors that received badges.
As shown in the figure, the counterfactual response trend (i.e., dotted line) is coefficient to the response trend in the control group. % add "in the" control group %non-donors.
Using that counterfactual response trend and response trend in donors (i.e., positive and negative coefficient values), we infer the effect of donation badges.
For instance, a negative coefficient value indicates a faster response time while a slower response time is indicated by a positive coefficient value.

%
%\todo{This figure shows the employment rates of two states at the time of before and after one of the state adopted an increase of its minimum wage (treatment).
%Assuming that employment trends would be the same in both states in the absence of the treatment, the counterfactual employment rate of the treatment state can be estimated as illustrated with a dotted line in the figure.
%The treatment effect can be derived from a difference between the actual employment rate and the counterfactual employment rate of the treatment state at the time of after treatment.}

% remove subsection
%\subsection{Quantile Differences in Differences}

To improve our results, the DID is extended to quantile differences in differences (QDID), to describe better the relationships at median and other quantiles (only median is studied in this paper because of the space limitation).
Although half of the reports got responses in one day, the average time is almost two months because of some outliers (i.e., the max value is more than four years).% add half of "the" reports 
%DID is analyzed with a standard regression model, which describes only the average relationship among variables.
%Considering response times in the study data, although half of reports got responses in one day, the average time is almost two months because of some outliers (the max value is more than four years).
% QDID enables us to discuss treatment effects at different quantiles.

%     Min.   1st Qu.    Median      Mean   3rd Qu.      Max. 
%   0.0003    0.0739    1.0510   56.6000   19.0200 1546.0000 
%\todo{
%Most of applied Econometrics is concerned with averages~\cite{angrist2008mostly}, and DID is also concerned with averages.
%Software engineering focuses more on median because most of the software engineering data are not normally distributed.
%}
% remove subsection
%\subsection{Propensity Score Matching}
Since DID depends on the common trends assumption~\cite{angrist2008mostly}, selecting a proper control group is necessary. %(no treatment) is necessary.
Matching is a statistical technique, for every member of donors, %change in dornors -> of donors %a treated group, 
to find a control member with similar observable characteristics, and is used to reduce selection bias by equating groups.
We use propensity score matching as it is a popular matching technique.

%Here is overview.
%Subsection explain part of process on this study.
\section{Approach}

\begin{figure*}[t]
\centering
  \includegraphics[width=\linewidth]{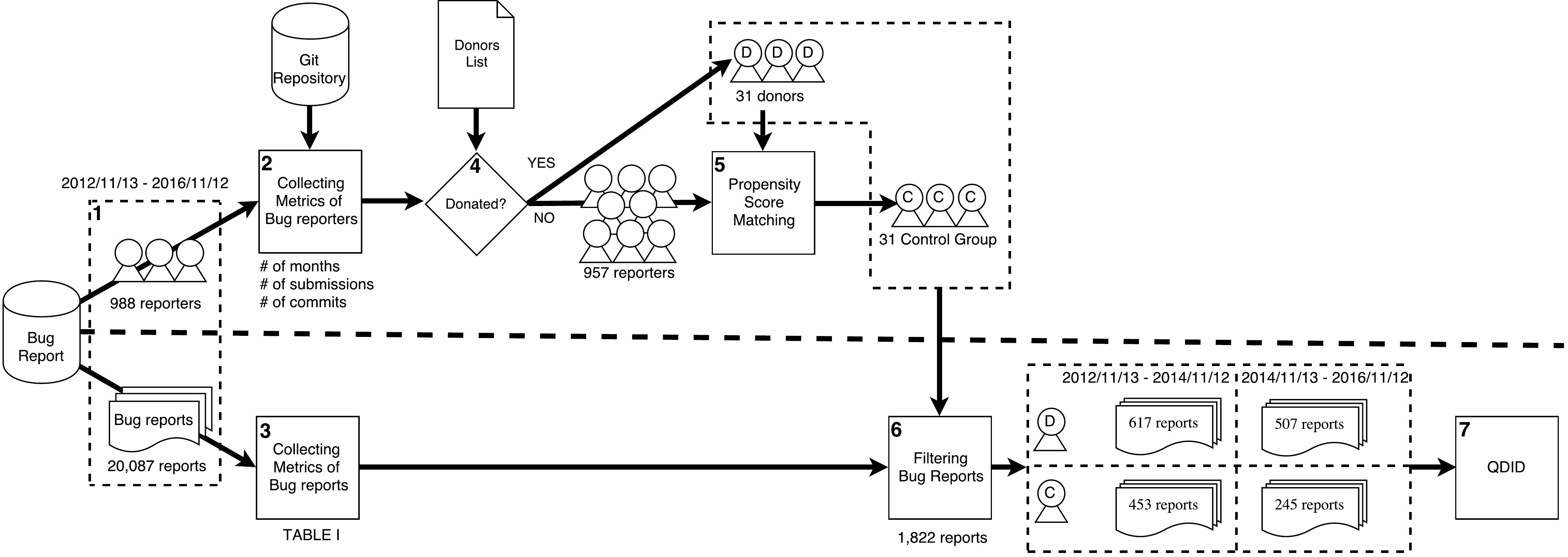}
  \caption{An overview of the analysis for data collection and causal inference.}
  \label{fig:overview}
\end{figure*}

Our analysis is composed of two phases as shown in Figure~\ref{fig:overview}. First, we select two groups of reporters, that is, a donors and a control group whose members have not donated (upper side), then two groups of bug reports in two time periods are identified, which are submitted by reporters in the above two reporter groups (lower side). We designed the analysis with the two phases instead of simply preparing two groups of bug reports (with and without donor badges) because the latter can cause a bias in selecting bug reports from specific reporters.

%Figure \ref{fig:overview} is the outline of this study. We designed two sections of experiments to evaluate of donation badges.
%The upper side of this figure shows the procedure of propensity score matching. The lower side of its shows the procedure of QDID (Quantile Difference in Differences). 
%The section of propensity score matching is the choice of the subject from the bug reporter. The section of QDID is performed evaluation of the donation badge. 
%Each of the collecting metrics is preprocessing part of propensity score matching and QDID. The filtering parts of QDID section performing of remove noise of bug report. 
%In the following subsections, we describe the detail of part of processing.

% remove subsection
%\subsection{Collecting Bug Reports and Metrics}

\textbf{Step 1: Bug report collection.}
From the discussion of proposing donation badges~\cite{bug434249}, we speculate that November 13, 2014 can be identified as the date of initially implementing donation badges.
So we first identify reporters who had submitted at least once in both periods, that is, two years earlier and later than the donation badge implementation.
We then collected bug reports submitted by the above reporters in a period between the date two years earlier and two years later than the implementation. Then, we removed bug reports whose first response comments were made by the same reporters or bug reports that were assigned to the original reporters.
Bug reports whose first responses had not come in three days were removed.
This is to exclude forgotten or intentionally postponed (weeks to years for the first responses) bug reports.
Furthermore, we analyze the impact at the hourly unit, by focusing on relatively promptly responded reports. 
Consequently, we are left with $60\%$ of reports after the removal.
%\todo{because they can be considered as outliers.}

%\begin{table*}[t]
%% increase table row spacing, adjust to taste
%\renewcommand{\arraystretch}{1.3}
% if using array.sty, it might be a good idea to tweak the value of
% \extrarowheight as needed to properly center the text within the cells
%\caption{Metrics of Propensity Score Matching}
%\label{tab:PSM_Metrics}
%\centering
%% Some packages, such as MDW tools, offer better commands for making tables
%% than the plain LaTeX2e tabular which is used here.
%\begin{tabular}{llrrrr}
%\hline
%\textbf{Metrics} & \textbf{Discription} & \textbf{Donor Ave.} & \textbf{Donor Med.} & \textbf{Control Ave.} & %\textbf{Control Med.}\\\hline
%Experience & month & 50.0 & 62 & 50.2 & 63\\
%\# of Before Submits & 2012-2014 & 28.2 & 12 & 20.2 & 5 \\
%\# of After submits & 2014-2016 & 22.8 & 6 & 10.3 & 4 \\
%\# of commits & before 2012 & 234.0 & 0 & 45.9 & 0 \\
%\hline
%\end{tabular}
%\end{table*}

\textbf{Step 2: Bug reporter metrics.}
Metrics of bug reporters are used in the propensity score matching (step 5).
The number of months worked in Bugzilla, the number of bug report submissions in the periods of before and after implementation, and the number of commits worked in Git repositories, are measured for all reporters.

%TABLE \ref{tab:PSM_Metrics} showed bug reporter metrics. The metrics of Experience is the reporter's the number of months of had submitted bug reports as of Nov. 2012.
%We analyze two terms. We define before term is from Nov. 2012 to Nov. 2014 and after term is from Nov. 2014 to Nov. 2016.
%A metric of number of Before Submits is the reporter's the number of submits in before term.
%A metric of number After Submits is the reporter's the number of submits in after term.
%A metric of number of commits is the count of the reporter's commit times.

\begin{table*}[!t]
%% increase table row spacing, adjust to taste
%\renewcommand{\arraystretch}{1.3}
% if using array.sty, it might be a good idea to tweak the value of
% \extrarowheight as needed to properly center the text within the cells
\caption{Explanatory variables and their effects in 5th decile (Quantile)}
\label{tab:QDID}
\centering
%% Some packages, such as MDW tools, offer better commands for making tables
%% than the plain LaTeX2e tabular which is used here.
\begin{tabular}{llrrrr}
\hline
\textbf{Metric} & \textbf{Description} & \textbf{Coeffs (Errors)} & \textbf{t value} & \textbf{Pr (\texttt{>|t|})} \\
\hline
(Intercept) & &2.237 (1.120) & 1.998 & 0.046 \\
Donor & Reporter is a donor or not & 2.259 (0.834) & 2.708 & 0.007\\
Period & Submitted time is before or after the badge introduction & -1.598 (1.227) & -1.302 & 0.193\\
\rowcolor{gray!20} Badge & It has a donation badge or not & -2.219 (1.061) & -2.092 & 0.037\\
Enhancement & Severity is enhancement or not & 0.668 (0.815) & 0.820 & 0.412\\
Windows & Issue is related to Windows or not & 1.119 (0.692) & 1.617 & 0.106\\
Linux & Issue is related to Linux or not & 0.675 (0.949) & 0.711 & 0.477\\
MacOS & Issue is related to MacOS or not & 0.793 (0.907) & 0.875 & 0.382\\
Component & Response days in median for the belonging components & 0.317 (0.198) & 1.603 & 0.109 \\
Community & \# of contributors in the belonging components  & 0.000 (0.001) & 0.105 & 0.917\\
Time & A numerical order of time in months & 0.152 (0.117) & 1.303 & 0.193\\
Relationship & \# of reports in which the reporter and the first responder have worked together & 0.005 (0.005) & 0.894 & 0.371\\
\hline
& & & \multicolumn{2}{r}{AIC = 4025, Pseudo $R^2$ = 0.011} \\
\end{tabular}
\end{table*}

%componentの上位から取ろうと思ったのですが，ほとんどが数件しかバグレポートがなかったので，100件以上あるcomponentのうち上位10と下位10をbean plotで描画しました．
%https://www.researchgate.net/post/Bean_Plot_Box_Plot-How_do_I_interpret_What_are_the_longer_lines_data_as_shown_in_attached_figure
%一番長い黒の線は中央値，点線は平均値，黒線は同じ数の分だけ長くなる．水色は密度推定
%\begin{figure*}[!t]
%\centering
%\subfloat[Top 10 components with earlier responses]{\includegraphics[width=.48\linewidth]{fig/top10.pdf}
%\label{fig:component_top10}}
%\hfil
%\subfloat[Top 10 components with later responses]{\includegraphics[width=.48\linewidth]{fig/worst10.pdf}
%\label{fig:component_worst10}}
%\caption{The distributions of developer response time in days per components. The beanplots of the top 10 earlier components (a) and the top 10 later components (b) are presented.}
%\label{fig:component}
%\end{figure*}

\textbf{Step 3: Bug report metrics.}
Metrics of bug reports (shown in Table~\ref{tab:QDID}) are used in QDID (step 7).
In addition to essential variables for DID (\textit{donor}, \textit{period}, and \textit{badge}), possible factors, which affect developer response time (in hours), were explored.
Several categorical variables were considered: severity metrics of seven severity levels (trivial to blocker and enhancement) and metrics related to operating systems.
It is reported that bugs with higher severity were fixed faster~\cite{Zhang:2013:PBT:2486788.2486931}, and operating systems where bugs were found were reported to influence bug fixing time~\cite{Zhang:2012:ESF:2420240.2420462}.
\textit{Component} metric measures the median of response time in component.
We found in our pilot study, that developer response times vary with different components. %Although, in some components, responses come within half a day in median, in other components, responses come later than 10 days in median.
\textit{Community} metric represents the size of contributors in community of the component. \textit{Time} metric is a numerical order of time in months, which was used in~\cite{Zhao:2017:ICI:3155562.3155575}. We prepared this metric to consider the impact of time. \textit{Relationship} metric mean to consider social and personal relationships between reporters and responders.
Ortu {\it et al.} found that emotional comments could influence fixing time~\cite{Ortu:2015:BMP:2820518.2820555}.
However, we did not add emotional factors because it is reported that existing sentiment analysis tools are not always applicable to software engineering domains~\cite{Jongeling:2017:NRU:3135854.3135907}.
From the above metrics, we select a subset of metrics based on the Akaike information criterion (AIC). Table~\ref{tab:QDID} shows such subset that derived the minimum AIC value.
For a response variable, response time was also measured as a hour interval between the moment a bug is reported and when it gets its first response comment.

%Bug report metrics are used in quantile difference in differences.
%TABLE \ref{tab:QDID} explain to metrics of bug reports. Metric of DID express the report with donation badge. We concern this metric whether or not it affect the comment time. We collect some reports information (e.g. Severity, Attachments, Description, Reporter). The Submit means which number of submit from Nov. 2012. This metric expresses reporters experience in analysis term.

% remove subsection
%\subsection{Selecting Bug Reporters}

\textbf{Step 4: Classifying donors or not.}
Names, the date of donations, and the amount of donations are summarized in a donor list.
From the above information, we can identify the periods of donation badge appearance in bug reports for each donor.
%We collect donors list. Donors list has donor's name, amount of donations, date of donations.
%We identify the report with donation badge from the reporter's date, their reporter whether or not in donor list, and their reporter whether or not in having Friend of Eclipse status from date of donations and amount of donations.
When duplicate names but different email addresses appear in bug reports, we removed them because we could not associate reporters and donor names uniquely. Consequently, 31 donors were identified.

\textbf{Step 5: Propensity Score Matching.} 
%Propensity score matching use reporter's information which create in step 2 and step 4. The aim of propensity score matching create similar reporter with donation badge from others because we would like to compare with donor and others but also the differences are only whether or not who donate.
We used the well-known nearest neighbor matching algorithm in propensity score matching.
From 957 reporters who had not donated,
31 reporters matched with the 31 donors are identified as a control group.
%we selected 31 reporters as a control group \fixed{by matching}. % add by matching

% remove subsection
%\subsection{Causal Inference}

%\textbf{Step 6: Study Bug Reports}
\textbf{Step 6: Filtering Bug Reports.}
%We combine result of step 5 and step 3. In step 5, we make analysis subjects from propensity score matching. we extract the analysis subjects report from step 3 data.
Bug reports submitted by 31 donors or 31 members in a control group are used in QDID.
As shown in Figure~\ref{fig:overview}, those bug reports are labeled with two time periods and two groups.

%\begin{figure}[!t]
%\centering
%  \includegraphics[width=\linewidth]{fig/quantile.pdf}
%  \caption{\fixed{The effect of the \textit{Badge} metric as a change in days (y-axis) along with quantile (x-axis).}}
%  \label{fig:DID}
%\end{figure}

%　使用したデータのコメントがつくまでの時間
% hours
%    Min.  1st Qu.   Median     Mean  3rd Qu.     Max. 
% 0.01806  0.61080  3.52300 13.07000 18.22000 71.99000

\textbf{Step 7: Quantile Difference in Differences.}
%We performed QDID by 10 quantiles as described in Section~\ref{sec:ci}.
Using the collected bug report metrics shown in Table~\ref{tab:QDID}, QDID is performed. We report its results of 5th decile.

\section{Results and Discussion}

%As seen in Figure \ref{fig:DID}, the effect of Badge metric shows consistently negative values (time decreasing) for each of the 10 quantiles.
Table~\ref{tab:QDID} shows coefficient values with p-values at 5th decile (median).
Only the Badge metric has a statistically significant positive effect (i.e.,　negative coefficient) on developer response time.
Its coefficient value indicating the estimated donation badge effect is about minus two hours. %The effects of the other metrics are not statistically significant.
%
%We discuss the estimated badge effect. The estimated badge effect is very effective the reason why the median of comment time of reports is 3.52. the 
Note that the effect size is relatively small (Pseudo $R^2 = 0.011$).
Since donation badges have been introduced only a few years ago, further analysis with longer histories is important.
%Although we measured several metrics that can affect response time, hidden factors may exist.

From the results, readers can infer three findings from Table \ref{tab:QDID}. (1) the response time is faster after contributors gained a donation badge.
This is evident by the negative coefficient of the \textit{Badge} metrics as shown in the table. 
Furthermore, (2) badges did not have negative effects: all responses to both donors and contributors became faster after badges were introduced.
This is evident by the negative coefficient of the \textit{Period} metrics. 
Finally, (3) the control group is not unfairly selected:
donors had longer response times compared to the control group. This is evident by the positive coefficient of the \textit{Donor} metrics.

%Considering coefficient
%values, although not statistically significant, (1) analyzed contributors
%ho did not have badges had lower response times
%(donor value is positive) and (2) response times decreased after
%the badge introduction (period value is negative).} %Considering coefficient values, although not statistically significant, (1) analyzed contributors who did not have badges had lower response times (donor value is positive) and (2) response times decreased after the badge introduction (period value is negative).}

Why do donation badges cause faster response times?
We assume that a donation badge works as a \textit{signal}, which is a perceivable indicator of hidden technical and social qualities~\cite{Tsay:2014:IST:2568225.2568315}.
In detail, donors are likely to contribute long-term instead of being one-time contributions,
Furthermore, they are more receptive and willing to help communities.
Thus, responders may infer and react to these qualities.

Considering the median of developer response time in 1,822 studied bug reports is 3.5 hours, decreasing time by two hours is not trivial.
This could be a practical rewarding effect for individual donors. For the organizers managing a developer ecosystem, donation badges has appeal because of the potential benefits at an inexpensive cost.
Since there is no project-specific metric nor assumption, we believe that our findings are not only limited to the current Eclipse project.

%memo
%Each black dot is the slope coefficient for the quantile indicated on the x axis. The red lines are the least squares estimate and its confidence interval. ref http://data.library.virginia.edu/getting-started-with-quantile-regression/

\section{Conclusion}

Applying a framework of causal inference from Econometrics, we investigated the causal effect of donation badges on Bugzilla, one of benefits for donors.
We estimated that donation badges decrease developer response time for bug reports by about two hours in median.
Our findings show the appearance of donation badges is appealing for both contributors and organizers.
Other OSS organizers are able to adopt this strategy to manage their developer ecosystems.

%\todo{framework for assessment}

% or


%\appendices

%\section{}

%\section*{Acknowledgment}
%This work has been supported by JSPS KAKENHI Grant Number 16H05857.


\ifCLASSOPTIONcaptionsoff
  \newpage
\fi

\end{document}